\documentclass[intlimits,twoside,a4paper]{article}

\usepackage[cp1251]{inputenc}

\usepackage[eqsecnum]{cmpj3}

\issue{2026}{29}{1}{13501}
\doinumber{10.5488/CMP.29.13501}
\title[Modelling SARS-CoV-2 epidemics]%
{Modelling SARS-CoV-2 epidemics via compartmental and cellular automaton \textit{SEIRS} model with temporal immunity and vaccination%
\thanks{The manuscript is dedicated to the memory of Prof. Stefan Soko{\l}owski, outstanding scientist, teacher, colleague and the friend}
}
\author[J. Ilnytskyi, T. Patsahan]
{J. Ilnytskyi\orcid{0000-0002-1868-5648}\refaddr{label1,label2}\thanks{Corresponding author: \email{iln@icmp.lviv.ua}},
T. Patsahan\orcid{0000-0002-7870-2219}\refaddr{label1,label2}
}
\addresses{
\addr{label1} Yukhnovskii Institute for Condensed Matter Physics of the National Academy of Sciences of Ukraine, 1~Svientsitskii~Str., 79011 Lviv, Ukraine
\addr{label2} Institute of Applied Mathematics and Fundamental Sciences, Lviv Polytechnic National University, 12~S.~Bandera~Str.,
79013 Lviv, Ukraine
}
\Keywords{compartmental models, cellular automata, epidemics}
\date{Received 19 January 2026; accepted 16 February 2026; published 30 March 2026}

\begin{document}

\maketitle

\begin{abstract}
We consider the \textit{SEIRS} epidemiology model with such features of the COVID-19 outbreak as: abundance of unidentified infected individuals, limited time of immunity and a possibility of vaccination. The control of the pandemic dynamics is possible by restricting the transmission rate, increasing identification and isolation rate of infected individuals, and via vaccination. For the compartmental version of this model, we found stable disease-free and endemic stationary states. The basic reproductive number is analysed with respect to balancing quarantine and vaccination measures. The positions and heights of the first peak of outbreak are obtained numerically and fitted to simple in usage algebraic forms. Lattice-based realization of this model is studied by means of the asynchronous cellular automaton algorithm. This permitted to study the effect of social distancing by varying the neighbourhood size of the model. The attempt is made to match the quarantine and vaccination effects.
      
\printkeywords
%
\end{abstract}




\section{\label{I}Introduction and the model}

It has been six and half years since the first outbreak of the pandemic caused by the SARS-CoV-2 virus took place in 2019, but its periodic local bursts \cite{Ge2022} continue to put a strain on the healthcare services and on various sectors of society. In more general terms, this pandemic put to a severe test the existing approaches to tackling newly emerging pathogens \cite{Chung2024}. At the early stages of a pandemic, the main goal was to stop its spread by all means, and the only measures available were strict individual hygiene and restrictive quarantine for social contacts termed as a lockdown. These measures were mostly effective in slowing the pandemic down at its early stage \cite{Flaxman2020, Hsiang2020, Bendavid2021, Bjornskov2021, Sharma2021, Perra2021}, but had an adverse effect on economy \cite{Yamaka2022, Wu2023}, mental well-being~\cite{Brooks2020, Singh2020, Ferwana2024}, and education process \cite{Aristovnik2020, DiPietro2023}. Various algorithms for gradual relaxation of a lockdown were suggested, see, e.g., reference~\cite{Block2020}. Development of a range of specialized vaccines~\cite{Li2022} opened up the possibility for a more balanced approach that combines both quarantine and repeated vaccination measures. However, immunization, acquired either in a natural way or via vaccination, turned out to be temporal due to  both physiological reasons and due to mutation of a virus itself \cite{Mallapaty2020, Hou2020, Goodman2020, Terry2020}. Most notable new variants of a virus were Alpha, Delta, Omicron, etc. \cite{VanNam2024, Tanneti2024, Irvem2025}. As a result, the pandemic turned into an decaying oscillatory process \cite{Pavlek2020, Huang2021, Odagaki2021, Ge2022, Jalal2024}. The merits of these oscillations, in terms of their time periodicity and evolution of respective amplitudes, is of much interest for predicting the most efficient measures that help to control the pandemic while causing a minimal impact on the society.

A wide range of mathematic models \cite{Erzen2020, Burch2024}, varied in their structure, features and assumptions, have been developed to reproduce and predict the development of a pandemic. They have evolved over time in response to the emerging evidence, e.g., gradually including the effects of virus mutation, vaccination and immunity loss. Currently, a broad range of  factors such as: travel restrictions \cite{Kucharski2020}, quarantine measures~\cite{Block2020, Rahimi2020}, mutation of a virus \cite{Yagan2021, Luo2024, AlThiabi2024, Ma2024, Saldana2024, Sarkar2024, Blyuss2025, Liu2025, Symons2025}, acquired immunity via vaccination \cite{Diagne2021, Kim2022, LaJoie2022, Adi2023, Jdid2024, Saleem2024, Cai2025, Juga2025, Blavatska2021}, and its gradual loss~\cite{Lopez2020} have been taken into account to various extent. However, ``the exact dynamics of waning immunity are still
uncertain, and are therefore incorporated into models using different approaches'', as remarked in reference~\cite{Burch2024}. The models are often parametrised  \cite{Pongkitivanichkul2020, Gotz2020} using available local or global statistical data from the WHO \cite{WHO_data} or local sources.

Most of the models are of the compartmental type~\cite{Brauer2008}, as pioneered by Kermack and McKend\-rick~\cite{Kermack1991}, yielding a set of ordinary differential equations (ODE). In this type of modelling, the population is split into a set of compartments, containing, e.g., susceptible, exposed, symptomatic and asymptomatic infective, recovered, vaccinated, isolated, hospitalized, etc.,  individuals with established transition rates between the compartments, and assuming strong mixing between individuals. The choice for the set of compartments is based upon the need to examine the role of particular process, namely: infection transmittance, identification of infected individuals, vaccination, virus mutation, etc. However, the attempt to take into account every possible process may overcomplicate the analysis of pandemic dynamics because of a large number of transition rates. 

Save for the simplest case of the \emph{SIS} model \cite{Kuhl2021}, the models of this type cannot be solved analytically. Typical workflow includes classification of stationary states of ODE, analysis of the basic reproduction number, and numeric solution for the pandemic dynamics at various combinations of transition rates. The aim of the current study is to construct rather minimalistic compartment model which takes into account a temporal nature of immunity, acquired either in a natural way or via vaccination, and undertaking the analysis according to the workflow outlined above. The basic reproductive number will be examined from the point of view of balancing the anti-pandemic measures. In this way, the study is a continuation of some previous works of ours \cite{Ilnytskyi2016, Ilnytskyi2017, Ilnytskyi2018, Ilnytskyi2021}. Despite a large number of previous works on the same topic, our study has a distinct feature in obtaining approximate algebraic forms for the positions and magnitudes of the first peak of a pandemic, as functions of a set of transition rates, which may have a practical use.
 
\begin{figure}[!ht]
\begin{center}
\includegraphics[clip,width=8cm,angle=0]{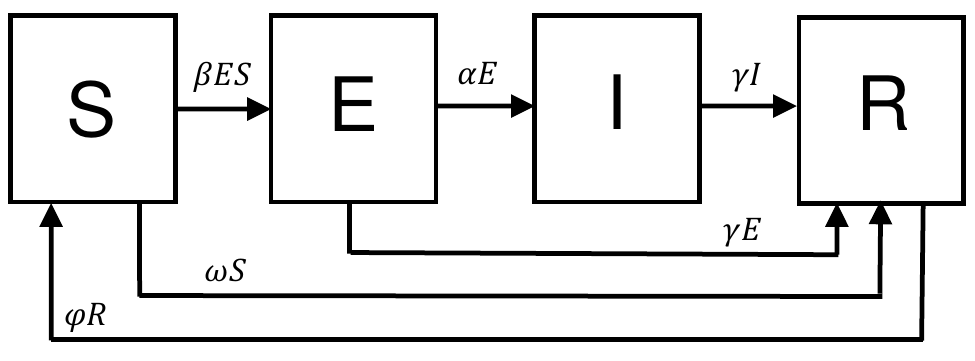}
\caption{\label{Model}The \textit{SEIRS} epidemiology model for the COVID-19 dissemination.}
\end{center}
\end{figure}

The minimalistic \textit{SEIRS} compartmental model, which incorporates the effects of vaccination and of a loss of immunity, is illustrated in figure~\ref{Model}. It contains four compartments, classified by the type of individuals they contain: susceptible to virus, $S$; unidentified infective, $E$; identified isolated infective, $I$; and recovered non-infective, $R$. The same letters, $S$, $E$, $I$, and $R$, denote respective fractions of individuals they contain.

\textit{Infecting} (transfer of an individual from $S$ to $E$ compartment) occurs with the rate of $\beta$. To keep the model simple, we do not introduce a separate compartment for exposed individuals \cite{Diagne2021, LaJoie2022, Jdid2024, Juga2025}. Therefore, the infecting rate, $\beta$, is affected by both social distancing, and by keeping sanitary norms, but these two factors cannot be separated in our model. Infected individual became instantly infective, as follows from the known facts of infectivity during the incubation period \cite{Sanche2020}. Infective individuals are \textit{identified} (transferred from $E$ to $I$), via testing, with a rate of $\alpha$. Again, not to overload a model with too many processes, we consider the case of an ideally disciplined society, where the identified infective individual becomes instantly isolated, either in home or in a hospital. The difference in curing in both cases is rather minor, save for fatal and severe forms. Therefore, we do not distinguish such patients, in contrast to some other models \cite{Diagne2021, Jdid2024}. Both unidentified and isolated infective individuals are supposed to \textit{recover}, acquire \textit{temporal immunity}, and loose their infectivity (transferred to $R$ compartment) with the same recovery rate of $\gamma$. \textit{Vaccination} provides a shortcut from the $S$ compartment straight into the $R$ one. We note, that vaccination in our model is explicit, and changed dynamically with the time evolution of the system, contrary to a static vaccination case \cite{Blavatska2021}, where the fraction of immune individuals is set constant. We neglect the vaccine activation period, and do not separate such factors as population coverage by vaccination and vaccine efficacy \cite{Jdid2024}, since both are combined into a single vaccination rate~$\omega$. In this way, all individuals residing in the $R$ compartment are assumed to be temporarily immunized. Upon \textit{loss of immunity} these are transferred back into the susceptible group $S$ with the loss of immunity rate $\varphi$. These ``returned'' individuals may be subject to reinfection \cite{Chen2024, Peghin2024, Fu2025}. This may occur due to the mutation of the SARS-CoV-2 virus, e.g., into Alpha, Delta, or Omicron variant, but this can be taken into account in our model only implicitly, via the increase of the loss of immunity rate of $\varphi$. The model also neglects the birth and death rates.

The rates $\beta$, $\alpha$, $\gamma$, $\varphi$ and $\omega$ can be related to the real life statistical data. If the population size is $N$ individuals, then the absolute number of susceptible, unidentified infective, identified isolated infective and recovered non-infective individuals are $N_S=SN$, $N_E=EN$, $N_I=IN$ and $N_R=RN$, respectively. Assuming a time unit equal to a single day, the absolute number of individuals infected per day is $\beta N_S (N_E/N)$.
The coefficient $\alpha$ reflects how widely the population is covered by appropriate medical testing and defines the number of newly identified infective individuals per day, $\alpha N_E$. Such type of data should be available from medical institutions. Recovery rate $\gamma$, is inversely proportional to the average period of infectivity, for which much data have been systemized in reference~\cite{Walsh2020}. There is some distribution of the results depending on many factors, but a reasonable average estimate of $14$ days is often cited \cite{canadaCOVID19Symptoms, Cheng2020, Arons2020, Singanayagam2020, LopezBernal2022}. This leads to the following estimate: $\gamma=1/14$. The loss of immunity rate, $\varphi$, inversely proportional to the time interval between (re)infections, is difficult to estimate as an unique average value~\cite{Townsend2021}. Though the early studies reported the time interval for the loss of immunity of 2--3 months \cite{dwCOVID19Longlasting}, 3--4 months \cite{aamcCOVIDRecently}, as well as long lasting immune memory \cite{Dan2021}, later works emphasize the role of vaccination, advanced age, the presence of comorbidities \cite{GmezGonzales2023}. The presence of chronic diseases, namely: ischemic and inflammatory heart disease, dysrhythmias, venous thromboembolism, cerebrovascular diseases are found as an important factor as well \cite{Tseng2025}. Therefore, one may face the need to fix the time interval to a certain value, e.g., $120$ days, yielding the estimate for $\varphi=1/120$. Finally, the vaccination rate, $\omega$, can be estimated from the absolute number of individuals vaccinated per day, $\omega N_S$, the information available from medical institutions.

The epidemiology model, shown in figure~\ref{Model} and discussed in detail above,  yields the following set of ODE with the normalization condition:
\begin{eqnarray}
&&\dot{S}  =  -\beta ES - \omega S + \varphi R \label{dSdt},\\
&&\dot{E}  =  \beta ES - (\gamma + \alpha) E \label{dEdt},\\
&&\dot{I}  =  \alpha E - \gamma I \label{dIdt},\\
&&\dot{R}  =  \gamma(E+I) + \omega S - \varphi R \label{dRdt},\\
&& S+E+I+R  = 1.\label{cond}
\end{eqnarray}
One can  clearly identify the effect of the presence of isolated $I$ and recovered $R$ individuals in this model by introducing cumulative fractions of all uninfected individuals, $S'=S+R$, and all infected ones, $E'=E+I$, and the normalization is: $S'+E'=1$. The equations set (\ref{dSdt})--(\ref{dRdt}) can be rewritten as a single equation for $E'$
\begin{equation}
\dot{E'} = \beta(E'-I)(S'-R) - \gamma E', \label{dEpdt}
\end{equation}
which has the same form as the one for the $SIS$ epidemiology model \cite{Ilnytskyi2016}, when $S$ and $I$ are substituted via $S'$ and $E'$, respectively, but with the reduced number of transmission acts. The latter are given by the product $(E'-I)(S'-R)$, where both infective and susceptible parties are lessened by $I$ and $R$, respectively. The equation (\ref{dEpdt}) is, obviously, not self-sufficient, as one needs to complement it by the equations providing the time evolution for the $I$ and $R$ fractions. To do so, one may attempt either to rewrite the equations set (\ref{dSdt})--(\ref{dRdt}) in terms of four variables $S'$, $E'$, $I$ and $R$, or to use certain approximations to express $E$ and $R$ fractions via $S'$ and $I'$ variables, similarly as suggested in reference~\cite{Ilnytskyi2021} for the \textit{SEIRS} model without the immunity and vaccination considered there.

The purpose of this study is to examine the stationary states and the time evolution of the \textit{SEIRS} model as defined in figure~\ref{Model} with the emphasis on the quarantine measures, the role of immunity loss and vaccination on pandemic dynamics. The study is of a general type with no direct link to particular country/region or statistical data of any sort. We, therefore, focus on the features and tendencies as predicted by this model rather than on practical recommendations that can be used straightaway. Section~\ref{II} contains the analysis of the stationary states (fixed points) for the model and analysis of their stability; in section \ref{III} we discuss early-time spread and the decay dynamics of the disease dissemination by combining numerical and approximate analytic tools; in section \ref{IV} we consider the equivalent cellular automaton model, section \ref{V} contains conclusions.

\section{Fixed points and their stability}\label {II}

The stationary states (fixed points) for the \textit{SEIRS} model are the solutions of the following equations set:
\begin{eqnarray}
&& -\beta ES - \omega S + \varphi R = 0, \label{Sfp}\\
&& \beta ES - (\gamma + \alpha) E = 0, \label{Efp}\\
&& \alpha E - \gamma I = 0, \label{Ifp}\\
&& \gamma(E+I) + \omega S - \varphi R = 0, \label{Rfp}\\
&& S + E + I + R = 1. \label{cond_fp}
\end{eqnarray}
We restrict our analysis to the case when both the loss of infectivity, $\gamma$, and the loss of immunity, $\varphi$, rates are constant and non-zero 
\begin{equation}\label{gp_nonzero}
\gamma=\mathrm{const}>0,\hspace{2em} \varphi=\mathrm{const}>0.
\end{equation}
This is logical, as the values for these two averaged over population, depend on the merits of the disease itself and cannot be easily affected.

\subsubsection*{Disease-free (DF) fixed point}

Hereafter we denote  all the fractions in the DF fixed point by the $\dagger$ superscript. The definition of the DF fixed point requires that $E^{\dagger}=I^{\dagger}=0$. It is easy to see from the equation~(\ref{Ifp}) that, if condition (\ref{gp_nonzero}) holds, then $I^{\dagger}=(\alpha/\gamma) E^{\dagger}$ and both these fractions turn into zero simultaneously. The set of equations is reduced to
\begin{eqnarray}
&& \omega S^{\dagger} - \varphi R^{\dagger} = 0, \label{dfeq1}\\
&& S^{\dagger} + R^{\dagger} = 1, \label{dfeq2}
\end{eqnarray}
yielding the solution:
\begin{equation}\label{DF_sol}
S^{\dagger}=\frac{\varphi}{\varphi+\omega}, \quad E^{\dagger}=I^{\dagger}=0, \quad R^{\dagger}=\frac{\omega}{\varphi+\omega}. 
\end{equation}
%

\subsubsection*{Endemic (EN) fixed point}

The fractions in the EN fixed point are denoted by the $^*$ superscript. The endemic fixed point is defined as such, that both $E^*>0$ and $I^*>0$. They are related via equation~(\ref{Ifp}) resulting in $I^*=(\alpha/\gamma) E^*$, with both $\gamma>0$ and $\alpha>0$, hence, a single condition $E^*>0$ is sufficient. In this case, both sides of equation~(\ref{Efp}) can be divided by $E^*$ resulting in a straightaway solution for $S^*=(\gamma+\alpha)/\beta$ for $S^*$. Now, the remaining equations in the set (\ref{Sfp})--(\ref{cond_fp}) are
\begin{eqnarray}
&& (\gamma+\alpha)E^* - \varphi R^* = - \omega S^*, \label{eneq1}\\
&& (\gamma+\alpha)E^* + \gamma R^* = \gamma (1-S^*). \label{eneq2}
\end{eqnarray}
The complete solution for the EN fixed point can be written in the following form
\begin{eqnarray}
S^* & = & \frac{\gamma+\alpha}{\beta}, \label{EN_sol_S}\\ 
E^* & = & \frac{\gamma\varphi}{(\gamma+\alpha)(\gamma+\varphi)}\left[1-S^*/S^{\dagger}\right], \label{EN_sol_E}\\
I^* & = & \frac{\alpha\varphi}{(\gamma+\alpha)(\gamma+\varphi)}\left[1-S^*/S^{\dagger}\right], \label{EN_sol_I}\\
R^* & = & \frac{\gamma}{\gamma+\varphi}\left[1-\left(1-\frac{\omega}{\gamma}\right)S^*\right] .\label{EN_sol_R}
\end{eqnarray}
Again, this solution exists only at $E^*>0$, which, according to equation~(\ref{EN_sol_E}) yields $S^*<S^\dagger$ for the~$S^*$ fraction. Following equation~(\ref{EN_sol_S}), one can see that, if the expression for $S^*={(\gamma+\alpha)}/{\beta}$ becomes equal or greater than $S^\dagger$, the crossover to the DF fixed point takes place. Using the expression for $S^\dagger$ (\ref{DF_sol}), this condition for the DF fixed point can be written as
\begin{equation}\label{R0}
R_0=\frac{\beta\varphi}{(\gamma+\alpha)(\varphi+\omega)} \leqslant 1,
\end{equation}
where $R_0$ has a meaning of the basic reproductive number. Let us note that in the limit case of $\alpha=\omega=0$ (no identification and no vaccination), the expression $R_0=\beta/\gamma$ for the \emph{SIS} model is retrieved (in this case the $E$ fraction of the current model serves as the $I$ fraction in the \emph{SIS} model). Taking into account the assumption (\ref{gp_nonzero}), the expression (\ref{R0}) indicates the way of bringing the basic reproductive number~$R_0$ down by means of both decrease of the transmission rate $\beta$ and by an increase of the identification $\alpha$ and vaccination $\omega$ rates. One can rewrite the condition $R_0 \leqslant 1$ for the DF state in any of the following three  forms
\begin{equation}\label{bc_ac_oc}
\beta\leqslant\beta_c=\frac{(\gamma+\alpha)(\varphi+\omega)}{\varphi},\quad \alpha\geqslant\alpha_c=\frac{\beta\varphi}{\varphi+\omega}-\gamma,\quad \omega\geqslant\omega_c=\frac{\beta\varphi}{\gamma+\alpha}-\varphi.
\end{equation}
One can interpret these inequalities as the requirements for the pandemic to be lowered. At fixed identification, $\alpha$, and vaccination, $\omega$, rates, this can be achieved only via quarantine measures that reduce infection rate below $\beta_c$. Similarly, at a fixed transmission, $\beta$, and vaccination, $\omega$, rates, this can be done only at sufficient coverage of population by tests, with identification  rate $\alpha\geqslant\alpha_c$. Finally, at fixed transmission,~$\beta$, and identification, $\alpha$, rates, the vaccination measures should be above the minimum rate of $\omega_c$.

\bigskip

As far as the pandemic can be lowered by a combination of such measures as isolation (lowering~$\beta$), wide coverage by tests (increasing $\alpha$) and vaccination (increasing $\omega$), the question arises concerning the most balanced approach in economical terms. For instance, a wider coverage of a population by tests may achieve the same level of the pandemic reduction as imposing a certain level of quarantine measures, but my means of less effect on the economy. The full differential of $R_0$
\begin{equation}\label{dR0}
\frac{1}{R_0}{\rd R_0}=\frac{1}{\beta}{\rd\beta} - \frac{1}{\gamma+\alpha}{\rd\alpha} - \frac{1}{\varphi+\omega}{\rd\omega}   
\end{equation}
provides the respective weights for the infinitesimal changes, $\rd\beta$, $\rd\alpha$ and $\rd\omega$, with which they enter the resulting infinitesimal change $\rd R_0/R_0$ of the basic reproductive number. If the particular model values for $\alpha$, $\beta$, $\gamma$, $\phi$, and $\omega$, as well as the respective financial burdens, $\rd\alpha$, $\rd\beta$,  and $\rd\omega$, can be estimated, then the most balanced strategy for lowering the pandemic can be envisaged.

Let us concentrate now on the linear stability analysis for both fixed points. To reduce the number of parameters, we eliminate the $R$ fraction by using the expression (\ref{cond}). Then, the equations set (\ref{dSdt})--(\ref{cond}) can be rewritten as
\begin{eqnarray}
\dot{S} & = & -\beta ES - \omega S + \varphi (1-S-E-I),\label{dSdt_st}\\
\dot{E} & = & \beta ES - (\gamma + \alpha) E, \label{dEdt_st}\\
\dot{I} & = & \alpha E - \gamma I. \label{dIdt_st}
\end{eqnarray}
The eigenvalues $\lambda$ of the Jacobian matrix $\mathbf{J}$ are given by the equation
\begin{equation}
\det \mathbf{J}=\left|
\begin{array}{ccc}
-\beta E-(\varphi+\omega)-\lambda & -\beta S-\varphi& -\varphi\\
\beta E & \beta S-(\gamma+\alpha)-\lambda & 0\\
0 & \alpha & -\gamma-\lambda
\end{array}
\right|=0.
\end{equation}
Replacing the determinant by its algebraic expression yields
\begin{equation}\label{lambda_eqs}
(\lambda+\gamma)(\lambda+\beta E+\varphi+\omega)(\lambda-\beta S+\gamma+\alpha) + \beta E \left[(\lambda+\gamma)(\beta S + \varphi) + \alpha\varphi\right] = 0.
\end{equation}
For the DF fixed point, we substitute $\{S,E\}$ by $\{S^\dagger,E^\dagger\}$, the expressions for the latter are given by equation~(\ref{DF_sol}). This simplifies the equation to
\begin{equation}\label{lambda_eqs_DF}
(\lambda+\gamma)(\lambda+\varphi+\omega)(\lambda-\beta S^\dagger+\gamma+\alpha) = 0.
\end{equation}
and provides the following roots: $\lambda^\dagger_1=-\gamma$, $\lambda^\dagger_2=-(\varphi+\omega)$, and $\lambda^\dagger_3=\beta S^\dagger - (\gamma+\alpha)=-(\gamma+\alpha)(1-R_0)$. Since $\gamma > 0$ and $\varphi+\omega > 0$, then both $\lambda^\dagger_1$ and $\lambda^\dagger_2$ are always negative. One has $R_0\leqslant1$ in the DF fixed point, hence, $\lambda^\dagger_3\leqslant 0$ there. Therefore, the linear stability analysis indicates the DF fixed point to be stable at $R_0<0$ (where all roots $\lambda_i$ are negative), but cannot determine its stability at $R_0=1$.

For the EN fixed point, we substitute $\{S,E\}$ by $\{S^*,E^*\}$, the expressions for the latter are given by equations~(\ref{EN_sol_S}) and (\ref{EN_sol_E}). Hence, one sees that $\beta S^*=\gamma+\alpha$, and the equation~(\ref{lambda_eqs}) takes the following form
\begin{equation}\label{lambda_eqs_EN}
\lambda(\lambda+\gamma)(\lambda+\beta E^*+\varphi+\omega) + \beta E^* \left[(\lambda+\gamma)(\gamma+\alpha+\varphi) + \alpha\varphi\right] = 0.
\end{equation}
It can be rewritten as
\begin{align}
\lambda^3 + \left[\gamma+\varphi+\omega+\beta E^*\right]\lambda^2 + \left[\gamma(\varphi+\omega)+(\gamma+\alpha+\gamma+\varphi)\beta E^*\right]\lambda 
+(\gamma+\alpha)(\gamma+\varphi)\beta E^* = 0.
 \label{lambda_eqs_EN_2}
\end{align}
We  examine it in a graphic way, similarly to the case of the \textit{SEIRS} model with no immunity \cite{Ilnytskyi2021}, hence skipping handbook details \cite{korn2013mathematical} here. There are three variable parameters, $\alpha$, $\beta$ and $\varphi$, therefore we  consider the discriminant $Q$ of the cubic equation (\ref{lambda_eqs_EN_2}) and its three roots, $\lambda_1$, $\lambda_2$ and $\lambda_3$, as the functions of $\alpha$ and $\beta$ at a fixed vaccination rate $\omega$.

\begin{figure}[!ht]
\begin{center}
\includegraphics[clip,width=13cm,angle=0]{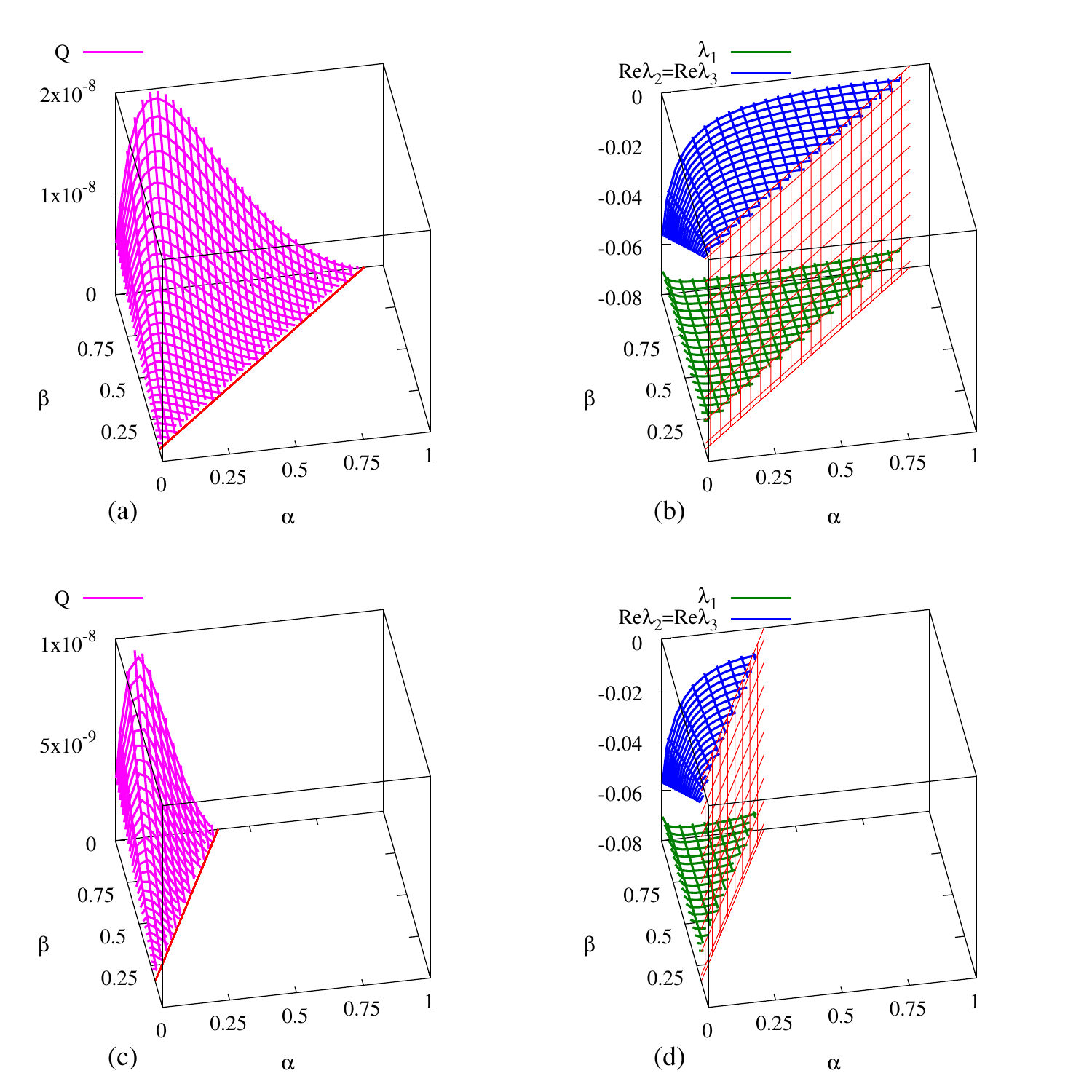}
\caption{\label{Q_lambda}(Colour online) Frame (a) shows the discriminant $Q$ of the cubic equation (\ref{lambda_eqs_EN_2}), frame (b) the real parts of its roots $\lambda_i$ for the no vaccination case, $\omega=0$. Frames (c) and (d) show the same respective characteristics for the vaccination rate of $\omega=0.01$.}
\end{center}
\end{figure}
The no vaccination case, $\omega=0$, is shown in the figure~\ref{Q_lambda}, frames (a) and (b), and it is considered first. All quantities of interest, $Q$ and $\lambda_i$, are shown only within the region $(\alpha,\beta)$, where $R_0>1$ and the EN fixed point (\ref{EN_sol_S})--(\ref{EN_sol_R}) is a valid solution. The crossover to the DF fixed point occurs along the line given by $R_0=1$, shown in red in the surface plot for $Q$. For the sake of clarity, it is translated along the $Z$-axis into the red dashed wall in the surface plots for $\lambda_i$. The surface representing $Q$ indicates positive values for the latter in the whole $R_0>1$ region, hence, the equation~(\ref{lambda_eqs_EN_2}) has one real, $\lambda_1$, and two complex, $\lambda_{2}$ and $\lambda_{3}$, roots. As it follows from the surface plots for $\lambda_1$ and for the real parts of $\lambda_{2}$ and $\lambda_{3}$, all three are negative in the whole $R_0>1$ region. This confirms the stability of the EN fixed point within this region, as examined graphically.

The case with the vaccination rate of $\omega=0.01$ is shown in the figure~\ref{Q_lambda}, frames (c) and (d). One observes the rotation of the $R_0$ line here and the  reduction of the $R_0>1$ area compared to the $\omega=0$ case. The surface plot for $Q$ keeps its general shape, but it is more ``jammed'' from the crossover line towards the $(\alpha=1$, $\beta=1)$ point. The values of $Q$ keep remaining positive in the whole $R_0>1$ region. In contrast to this, the surfaces for $\lambda_1$ and for the real parts of $\lambda_{2}$ and $\lambda_{3}$, appear to be the same as in the case of $\omega=0$, but are cut now at a new position of the crossover line $R_0=1$. All three are negative within the $R_0>1$ region. The same holds true upon the further increase of the vaccination rate $\omega$ (not shown), until the $R_0>1$ region moves out of the square defined by the $0<\alpha<1$ and $0<\beta< 1$ boundaries. As a result, we conclude that the EN fixed point is stable within the $R_0>1$ region at all vaccination rates $\omega$.
    
\section{Numerical integration of differential equations and the approximate expressions for the first pandemic peak}\label {III}

Here we employ a numerical integration of equations~(\ref{dSdt})--(\ref{dRdt}), performed via the second-order integrator
\begin{equation}
X(t+\Delta t)=X(t) + \dot{X}(t)\Delta t + \frac{1}{2}\ddot{X}(t)\Delta t^2\label{integr}
\end{equation}
for each set of fractions, $X=\{S,E,I,R\}$, applied iteratively with the time step $\Delta t$ assumed to be equal to a single day. At time instance $t=0$, the system is characterized by the initial fraction of unidentified infected individuals $E(0)=E_0$, assumed to be brought from outside, and the other fractions are: $S(0)=1-E_0$, and $I(0)=R(0)=0$. Different values of $E_0$ are examined. The equations are coupled, as far as both the first derivatives $\dot{X}$, given by equations~(\ref{dSdt})--(\ref{dRdt}), and the second derivatives
%
\begin{eqnarray}
\ddot{S} & = & -\beta(S\dot{E} + E\dot{S}) - \omega \dot{S} + \varphi \dot{R}, \label{ddSdt}\\
\ddot{E} & = & \beta(S\dot{E}+ E\dot{S}) - (\gamma + \alpha) \dot{E}, \label{ddEdt}\\
\ddot{I} & = & \alpha \dot{E} - \gamma \dot{I}, \label{ddIdt}\\
\ddot{R} & = & \gamma (\dot{E}+\dot{I}) + \omega \dot{S} - \varphi \dot{R}, \label{ddRdt}
\end{eqnarray}
at the time instance $t$ depend on all variables, $S$, $E$, $I$ and $R$ at the same instance $t$. The source of infection is the $E$ fraction, because in the current model, depicted in figure~\ref{Model}, we consider an ideal case of instant isolation of identified infective individuals into a group $I$. On the other hand, part of the latter are isolated in the hospitals and at home, therefore $I$ reflects the load put on the health system. Therefore, we concentrate on the time evolutions, $E(t)$ and $I(t)$, of these two fractions. These are examined at various initial values of $E_0$ and at various $\beta$, $\alpha$ and $\omega$ rates.

\subsection{The no vaccination case, $\omega=0$}

\begin{figure}[!b]
\begin{center}
\includegraphics[clip,width=14cm,angle=0]{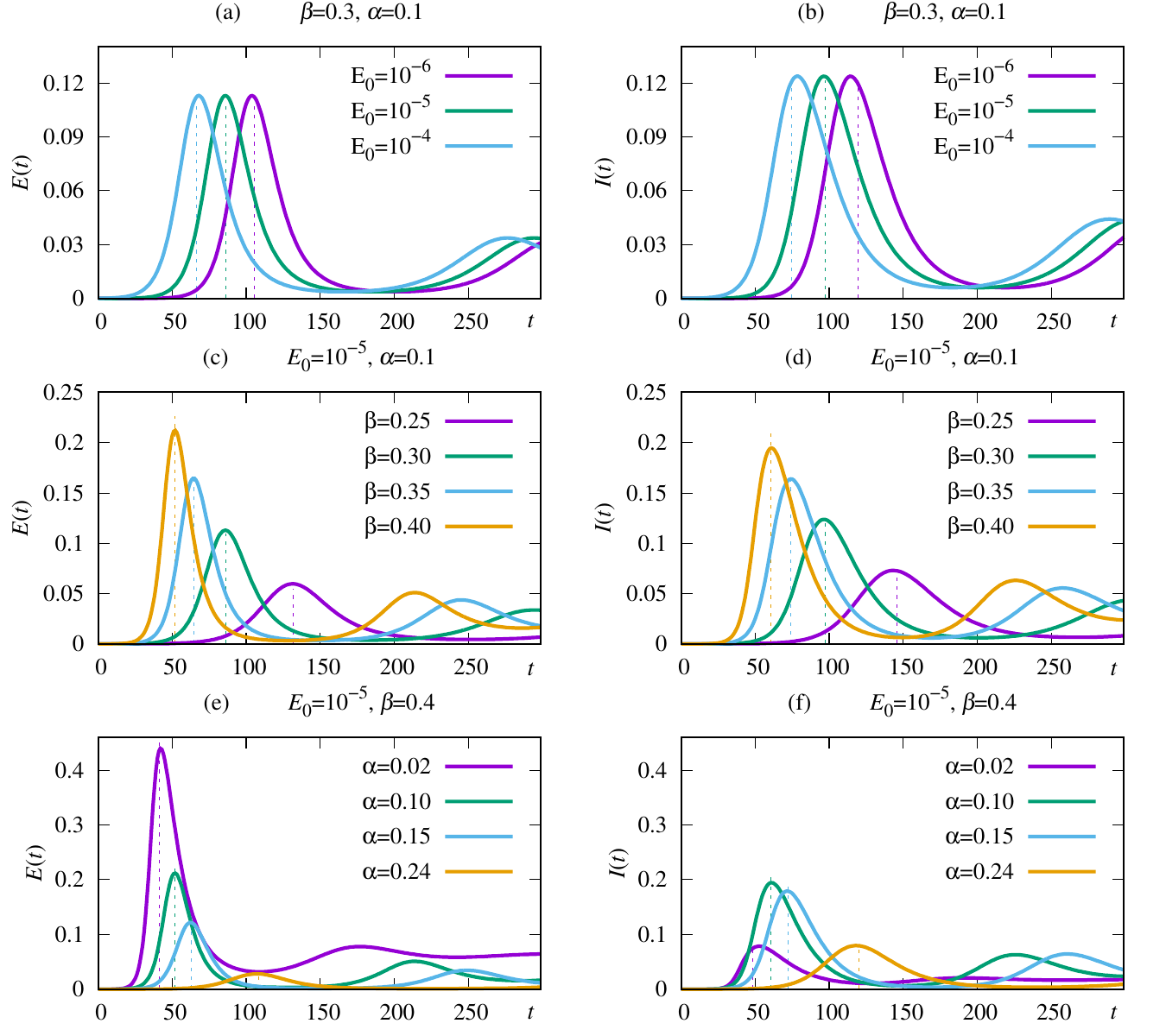}
\caption{\label{novacc_EI_evol}(Colour online) Numeric results for the time evolutions of the unidentified infected, $E(t)$, and isolated infective, $I(t)$, fractions at various model parameters. The no vaccination case, $\omega=0$, is shown. Frames (a) and (b) demonstrate the effect of variation of the initial value $E_0$ at fixed $\beta$ and $\alpha$; (c) and (d) show the effect of the contact rate $\beta$ at fixed $E_0$ and $\alpha$; (e) and (f) demonstrate the same for the identification rate $\alpha$ at fixed $E_0$ and $\beta$. Dashed vertical lines show approximate positions and heights of the first peak of a pandemic, and are the results of approximate analytic expressions, see explanation further below in the text.}
\end{center}
\end{figure}
We first discuss the no vaccination case, $\omega=0$. Contrary to the \textit{SEIRS} model with no immunity~\cite{Ilnytskyi2021}, we observe an oscillatory behaviour for $E(t)$ and $I(t)$ for most combinations of $\beta$ and $\alpha$ rates being considered. Both plots in the figures~\ref{novacc_EI_evol}(a) and (b) indicate that, at fixed $\beta$ and $\alpha$, the decrease of $E_0$ does not affect the first peak heights, $E_{\mathrm{max}}$ and $I_{\mathrm{max}}$, for both fractions, but shifts their respective positions, $t_{\mathrm{max},E}$ and $t_{\mathrm{max},I}$, towards the later times. The amount of a shift is found to be proportional to $-\log E_0$. The plots displayed in the figures~\ref{novacc_EI_evol}(c) and (d) show that a decrease of the contact rate $\beta$ decreases both $E_{\mathrm{max}}$ and $I_{\mathrm{max}}$ and shifts $t_{\mathrm{max},E}$ and $t_{\mathrm{max},I}$. The same effect is achieved by an increase of the $\alpha$ rate, as shown in figures~\ref{novacc_EI_evol}(e) and (f).

\begin{figure}[!ht]
\begin{center}
\includegraphics[clip,width=13cm,angle=0]{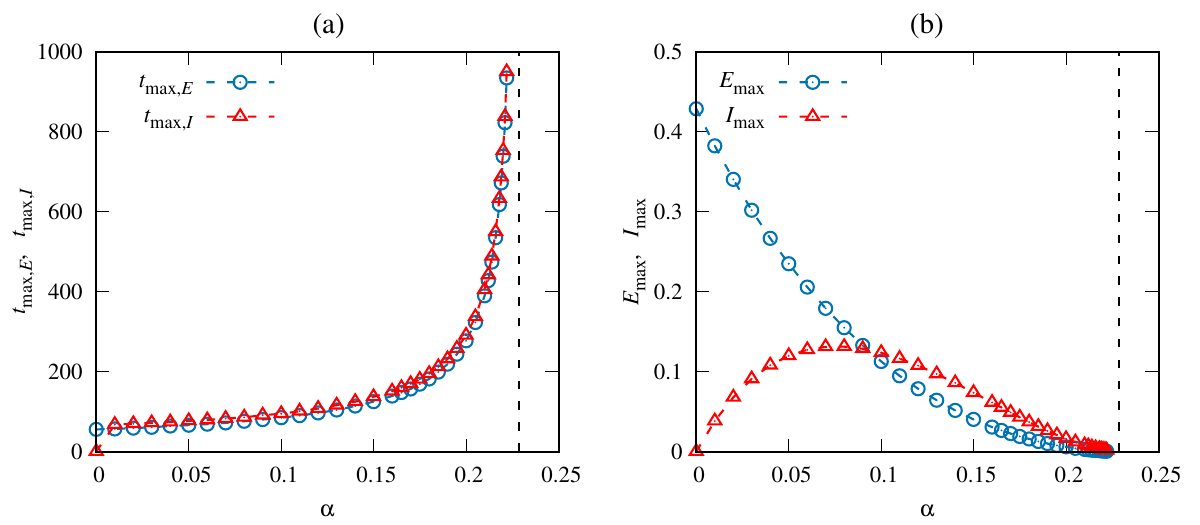}
\caption{\label{tmax_EImax_ex}(Colour online) Example for the $t_{\mathrm{max},E}$, $t_{\mathrm{max},I}$ (a), as well as $E_{\mathrm{max}}$ and $I_{\mathrm{max}}$ (b) as the functions of the identification rate $\alpha$ at a fixed transmission rate $\beta=0.3$ and initial condition $E_0=10^{-5}$. Vertical dashed lines provide the position of the critical identification rate $\alpha_c$, given by equation~(\ref{bc_ac_oc}).}
\end{center}
\end{figure}
Both the first peak heights, $E_{\mathrm{max}}$ and $I_{\mathrm{max}}$, and their respective positions, $t_{\mathrm{max},E}$ and $t_{\mathrm{max},I}$, are of great practical interest for predicting of the pandemic development. It is, however, impractical in solving equations~(\ref{dSdt})--(\ref{dRdt}) numerically at each required parameters set for this purpose, and one would improve the predictive practicality of the modelling approach by suggesting simple approximate expressions instead. To this end, we examined the general shape for all properties of interest, $t_{\mathrm{max},E}$, $t_{\mathrm{max},I}$, $E_{\mathrm{max}}$ and~$I_{\mathrm{max}}$, as the functions of the identification rate $\alpha$ at various fixed transmission rate $\beta$ and initial conditions given by $E_0$. We found that both peak positions, $t_{\mathrm{max},E}$ and $t_{\mathrm{max},I}$, diverge as $\alpha$ approaches~$\alpha_c$, where the latter is defined in equation~(\ref{bc_ac_oc}), see figure~\ref{tmax_EImax_ex}(a). Both peak heights, $E_{\mathrm{max}}$ and $I_{\mathrm{max}}$, decay to zero as $\alpha$ approaches $\alpha_c$, see figure~\ref{tmax_EImax_ex}(b). These observations, alongside with the one from figure~\ref{novacc_EI_evol}, that both $t_{\mathrm{max},E}$ and $t_{\mathrm{max},I}$ are proportional to $-\log E_0$ and $E_{\mathrm{max}}$ and $I_{\mathrm{max}}$ are independent of $E_0$, led to suggesting the scaling expressions of the following form
\begin{eqnarray}
t_{\mathrm{max},E}(\alpha,\beta,E_0) &=& -\log(E_0)A(\beta)\left[1-\frac{\alpha}{\alpha_c(\beta)}\right]^{-v(\beta,E_0)},\label{tmaxE}\\
t_{\mathrm{max},I}(\alpha,\beta,E_0) &=& -\log(E_0)B(\beta)\left[1-\frac{\alpha}{\alpha_c(\beta)}\right]^{-w(\beta,E_0)},\label{tmaxI}\\
E_{\mathrm{max}}(\alpha,\beta) &=& C(\beta)\left[1-\frac{\alpha}{\alpha_c(\beta)}\right]^{p(\beta)},\label{Emax}\\
I_{\mathrm{max}}(\alpha,\beta) &=& D(\beta)\frac{\alpha}{\alpha_c(\beta)}\left[1-\frac{\alpha}{\alpha_c(\beta)}\right]^{q(\beta)},\label{Imax}
\end{eqnarray}
where
\begin{equation}
	\alpha_c(\beta) = \beta - \gamma.\label{alphac}
\end{equation}
Functional forms for all unknown coefficients, $A(\beta)$, $B(\beta)$, $C(\beta)$ and $D(\beta)$, and related exponents, $v(\beta,E_0)$, $w(\beta,E_0)$, $p(\beta)$ and $q(\beta)$, are obtained by fitting the data obtained by the numeric solution given by expression (\ref{integr}). Let us remark, that, contrary to the theory of phase transitions \cite{PhTr}, where one seeks the universal critical exponents in the vicinity of a critical point (in our case, when $\alpha\to\alpha_c$), we opted here for the so-called effective critical exponents that are capable of approximating the properties of interest in a wider range of the $\alpha$ rate. Therefore, the exponents $v$, $w$, $p$ and $q$ are the functions of either~$\beta$ or both $\beta$ and $E_0$.

\begin{figure}[!ht]
\begin{center}
\includegraphics[clip,width=14cm,angle=0]{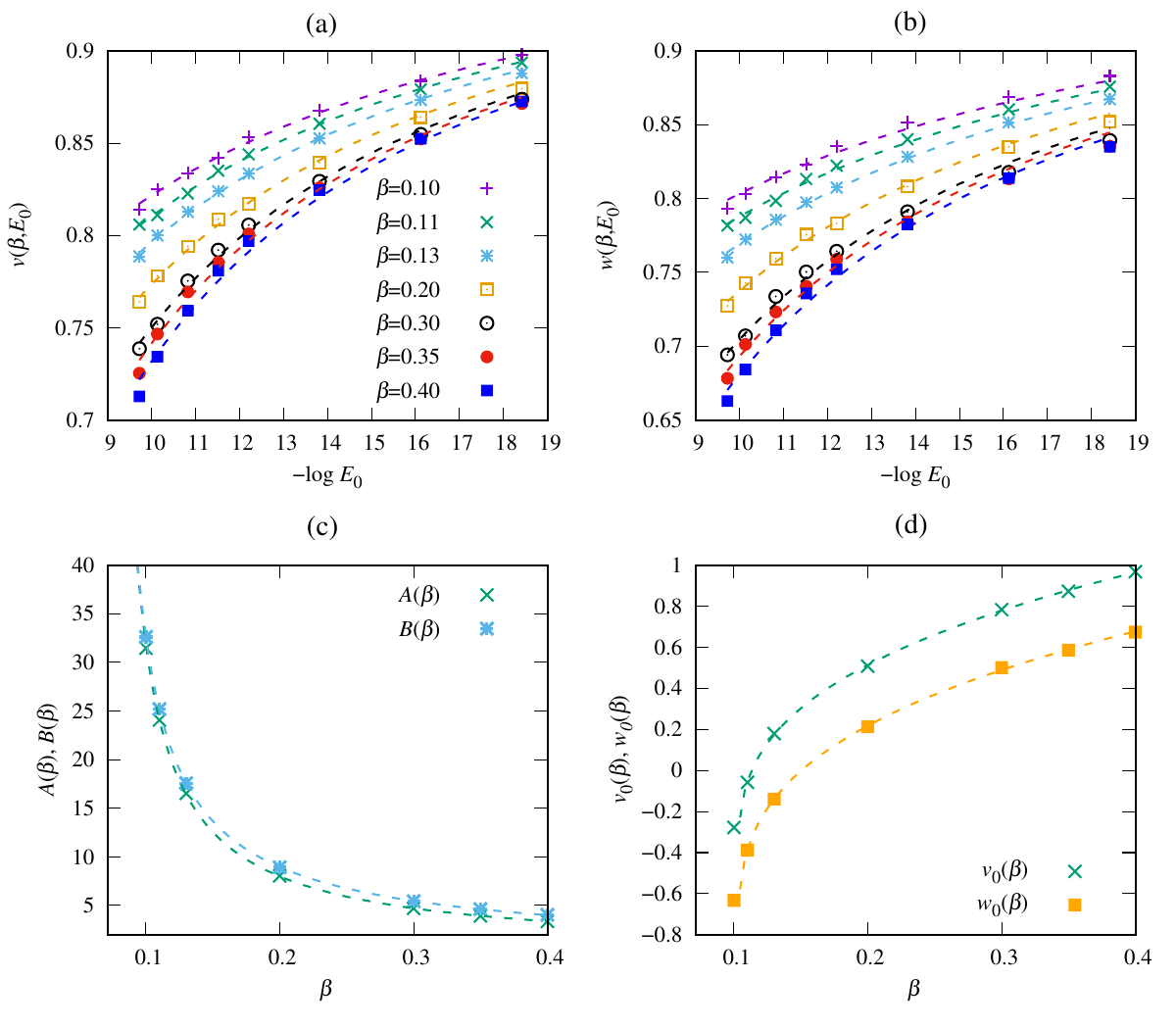}
\caption{\label{tmax_coeff}(Colour online) The results of fitting of $t_{\mathrm{max},E}(\alpha,\beta,E_0)$ (\ref{tmaxE}) and $t_{\mathrm{max},I}(\alpha,\beta,E_0)$ (\ref{tmaxI}). (a)~shows the exponent $v(\beta,E_0)$ as the function of $E_0$ at fixed $\beta$. (b)~shows the same for the exponent 
	$w(\beta,E_0)$.  (c)~shows the amplitudes $A(\beta)$ and $B(\beta)$.
	(d)~shows the exponents 
	$v_0(\beta)$ and $w_0(\beta)$ entering expressions (\ref{v0}) and (\ref{w0}), respectively.}
\end{center}
\end{figure}
The peak height coefficients and exponents, as the functions of their respective arguments, are shown in figure~\ref{tmax_coeff}. They can be fitted by the following expressions
\begin{eqnarray}
A(\beta)&=&1.22/(\beta-\gamma)^{0.92}\label{A},\\
v(\beta,E_0) &=& \frac{2}{\piup}\arctan[-v_0(\beta) + 0.32(-\log E_0)],\label{v}\\
v_0(\beta)&=&-0.40+2.11(\beta-0.104)^{0.36}\label{v0},\\
B(\beta)&=&1.54/(\beta-\gamma)^{0.86}\label{B},\\
w(\beta,E_0) &=& \frac{2}{\piup}\arctan[-w_0(\beta) + 0.25(-\log E_0)],\label{w}\\
w_0(\beta)&=&-0.81+2.22(\beta-0.104)^{0.33}\label{w0}.
\end{eqnarray}
\begin{figure}[!ht]
\begin{center}
\includegraphics[clip,width=13cm,angle=0]{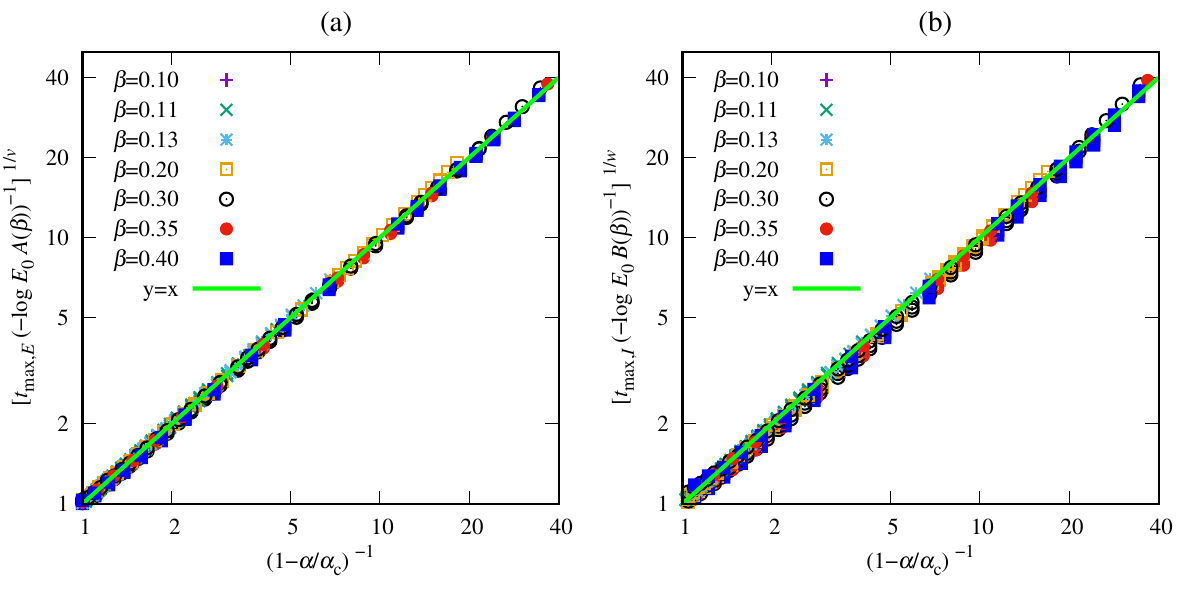}
\caption{\label{tmax_scl}(Colour online) Accuracy check for the scaling expressions (\ref{tmaxE}) (a) and (\ref{tmaxI}) (b). A wide range of transmission rates, $\beta\in[0.1,0.4]$, is displayed, where for each $\beta$ three different initial conditions, $E_0=10^{-7}$, $10^{-6}$ and $10^{-5}$ are studied and displayed in the plot.}
\end{center}
\end{figure}
As an accuracy check, we display the scaling plots for the combinations $\left[t_{\mathrm{max},E}/(-\log(E_0)A)\right]^{1/v(\beta,E_0)}$ and $\left[t_{\mathrm{max},I}/(-\log(E_0)B\right]^{1/w(\beta,E_0)}$ vs $(1-\alpha/\alpha_c)^{-1}$, obtained for the set of values of $\beta$ and $E_0$, see figure~\ref{tmax_scl}. The data, obtained by means of numeric integration, are found to  very closely follow the expected $y(x)=x$ dependence.

\begin{figure}[!ht]
\begin{center}
\includegraphics[clip,width=13cm,angle=0]{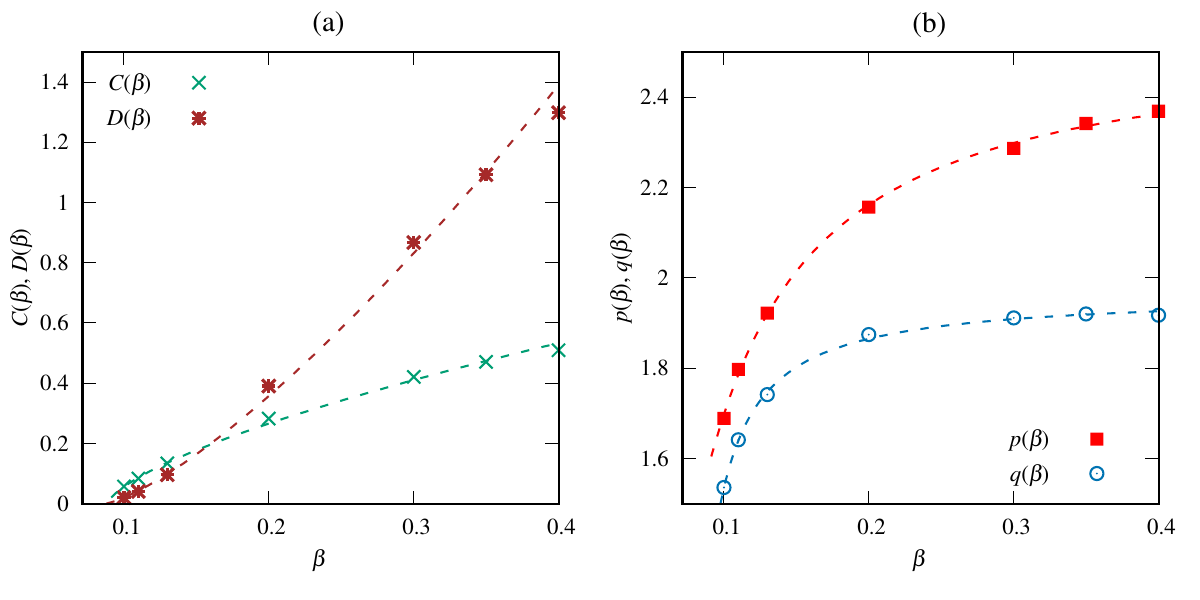}
\caption{\label{EImax_coeff}(Colour online) Coefficients $C(\beta)$ and $D(\beta)$ (a) and the exponents, $p(\beta,E_0)$ and $q(\beta,E_0)$ (b), for the scaling expressions (\ref{Emax}) and (\ref{Imax}).}
\end{center}
\end{figure}
\begin{figure}[!ht]
\begin{center}
\includegraphics[clip,width=13cm,angle=0]{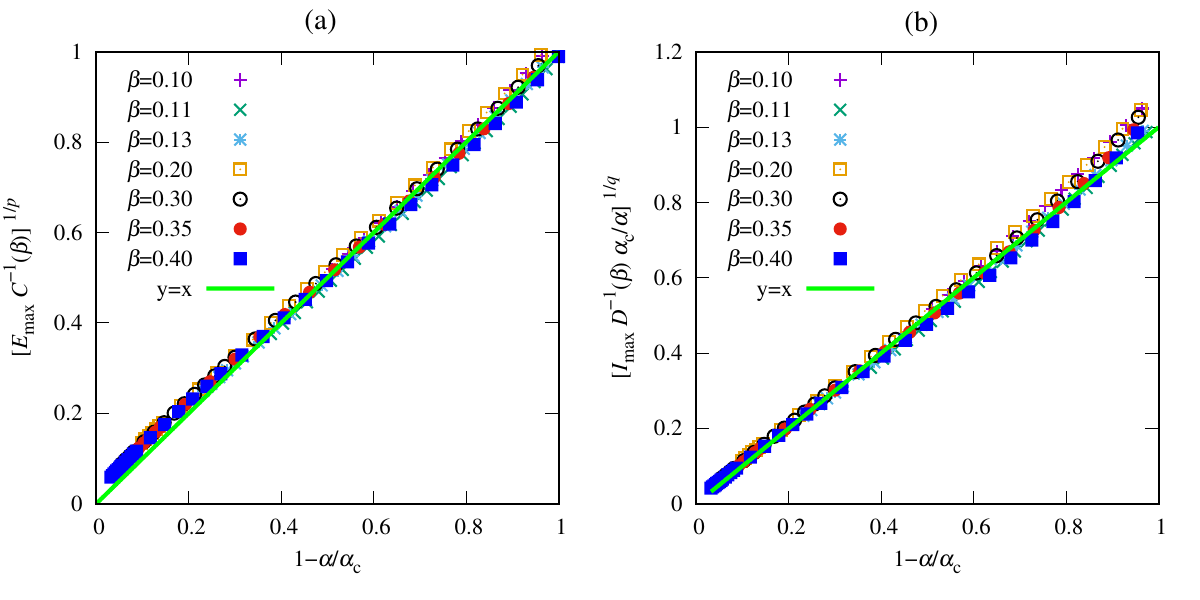}
\caption{\label{EImax_scl}(Colour online) The same as in figure~\ref{tmax_scl} but for the scaling expressions (\ref{Emax}) (a) and (\ref{Imax}) (b).}
\end{center}
\end{figure}

In a similar way, the peak positions coefficients and exponents, as the functions of their respective arguments, are shown in figure~\ref{EImax_coeff}. They can be fitted by the following expressions
\begin{eqnarray}
C(\beta)&=&1.18/(\beta-0.089)^{0.68},\label{C}\\
p(\beta) &=& 1.62\arctan(-0.70+24.46\beta),\label{p}\\
D(\beta)&=&6.53/(\beta-0.089)^{1.33},\label{D}\\
q(\beta) &=& 1.25\arctan(-6.89+96.82\beta).\label{q}
\end{eqnarray}
Again, as an accuracy check, we display the scaling plots for the combinations $[E_{\mathrm{max}}/C]^{p(\beta)}$ and $[I_{\mathrm{max}}\alpha_c/\alpha/D]^{q(\beta)}$ vs $1-\alpha/\alpha_c$ in figure~\ref{EImax_scl} and found these to follow the $y(x)=x$ dependence reasonably well, too.

Hence, in this section we performed a numerical integration (\ref{integr}) of the system of ODE (\ref{dSdt})--(\ref{dRdt}) focusing at the time evolutions of the fractions of unidentified, $E(t)$, and identified isolated, $I(t)$, individuals. Infection starts from the non-zero initial fraction $E_0=E(0)$, and three values for these were examined, $E_0=10^{-6}$, $10^{-5}$, and $10^{-4}$. Both the recovery rate, $\gamma=1/14$, and the loss of immunity rate, $\varphi=1/120$, were fixed, as discussed in section~\ref{I}. Time evolutions, shown in figure~\ref{novacc_EI_evol}, demonstrate a decaying oscillating behaviour for both $E(t)$ and $I(t)$ within a wide range of $\alpha$ and $\beta$ (the cases of $0.02\leqslant\alpha\leqslant 0.24$ and $0.25\leqslant\beta\leqslant 0.4$ are shown). We examined the position and the magnitude for the first peak in each case, and found that its position for both fractions displays a critical-like behaviour when approaching $\alpha_c$. Hence, by applying some strong measures in identification and isolation of infected individuals, characterized by $\alpha_c$, one may postpone the first pandemic peak to an infinite time. Based on this type of dependency, we suggested scaling-like approximate algebraic expressions (\ref{tmaxE})--(\ref{alphac}) for the positions and magnitudes of the first peak as the functions of the model rates. These have been checked to reproduce their counterparts obtained via numeric integration with a high accuracy.

\subsection{The vaccination case, $\omega>0$}

\begin{figure}[!ht]
\begin{center}
\includegraphics[clip,width=14cm,angle=0]{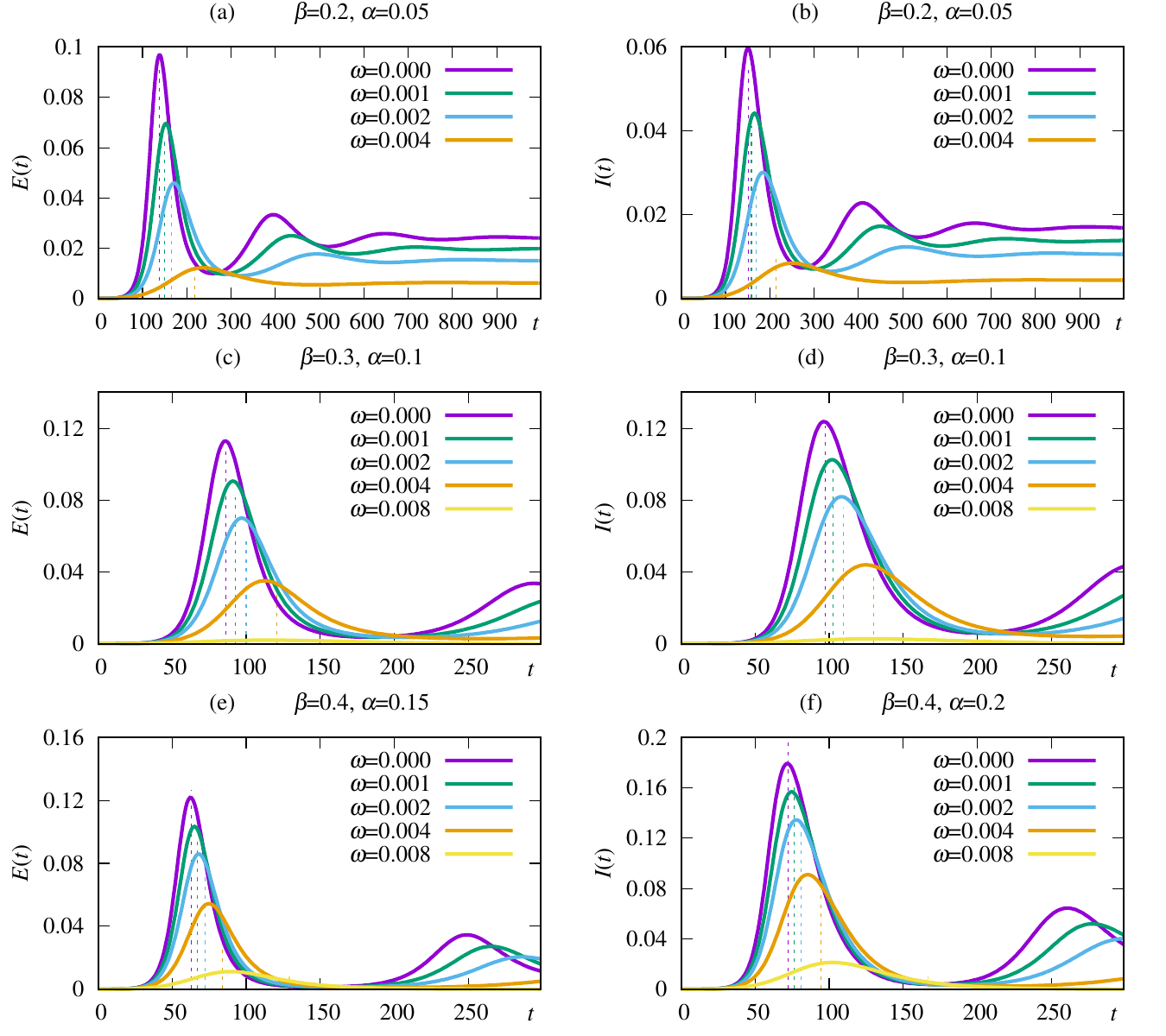}
\caption{\label{vacc_EI_evol}(Colour online) The effect of vaccination rate $\omega$ on the time evolution of the fractions $E(t)$ and~$I(t)$ at various values for the $\alpha$ and $\beta$ rates, as designated in respective frames (a)--(f) in the figure. Dashed vertical lines show approximate positions and heights of the first peak and are the results of approximate analytic expressions, see explanation in the text.}
\end{center}
\end{figure}
In this section we take into account vaccination of population undertaken with variable rate $\omega>0$. Vaccination leads to two noticeable effects. The first one is the shift of respective first peaks for $E(t)$ and $I(t)$ fractions to longer times; whereas the second one is an essential reduction of their respective magnitudes, see figure~\ref{vacc_EI_evol}. The approximate expressions for these read
\begin{eqnarray}
t_{\mathrm{max},E}(\alpha,\beta,E_0,\omega) &=& -\log(E_0)A(\beta)\tau_E(\beta,\omega)\left[1-\frac{\alpha}{\alpha^E_c(\beta,\omega)}\right]^{-v(\beta,E_0)},\label{tmaxEv}\\
t_{\mathrm{max},I}(\alpha,\beta,E_0,\omega) &=& -\log(E_0)B(\beta)\tau_I(\beta,\omega)\left[1-\frac{\alpha}{\alpha^I_c(\beta,\omega)}\right]^{-w(\beta,E_0)},\label{tmaxIv}\\
E_{\mathrm{max}}(\alpha,\beta,\omega) &=& C(\beta)\mu_E(\beta,\omega)\left[1-\frac{\alpha}{\alpha^E_c(\beta,\omega)}\right]^{p(\beta)},\label{Emaxv}\\
I_{\mathrm{max}}(\alpha,\beta,\omega) &=& D(\beta)\mu_I(\beta,\omega)\frac{\alpha}{\alpha^I_c(\beta,\omega)}\left[1-\frac{\alpha}{\alpha^I_c(\beta,\omega)}\right]^{q(\beta)},\label{Imaxv}
\end{eqnarray}
where
\begin{equation}
	\alpha^E_c(\beta,\omega) = \left(\frac{\beta\varphi}{\omega+\varphi} - \gamma\right)\exp(45\omega)\label{alphaEc}
\end{equation}
and
\begin{equation}
	\alpha^I_c(\beta,\omega) = \left(\frac{\beta\varphi}{\omega+\varphi} - \gamma\right)\exp(35\omega).\label{alphaIc}
\end{equation}
The expressions for $A(\beta)$, $v(\beta,E_0)$, $B(\beta)$, $w(\beta,E_0)$, $C(\beta)$, $p(\beta)$, $D(\beta)$ and $q(\beta)$ have been obtained above, see equations~(\ref{A})--(\ref{q}). The critical value for $\alpha$ now depends on $\omega$  and the expressions for $\alpha^E_c(\beta,\omega)$ and $\alpha^I_c(\beta,\omega)$ differ by the exponent argument prefactor only. The no vaccination limit case, $\alpha^E_c(\beta,0)=\alpha^I_c(\beta,0)=\alpha_c(\beta)$, where $\alpha_c(\beta)$ is given by equation~(\ref{alphac}), holds. On top of the dependence of the critical value on $\omega$, both peak positions and heights in equations~(\ref{tmaxEv})--(\ref{Imaxv}) acquires additional, $\omega$-dependent, multipliers,
\begin{eqnarray}
\tau_E(\beta,\omega) &=& \exp[-5\omega(\beta-\gamma)],\label{tauE}\\ 
\tau_I(\beta,\omega) &=& \exp\left(-\frac{5\omega}{\beta-\gamma}\right),\label{tauI}\\ 
\mu_E(\beta,\omega) &=& \exp\left(-\frac{19\omega}{\beta-\gamma}\right),\label{muE}\\ 
\mu_I(\beta,\omega) &=& \exp\left(-\frac{32\omega}{\beta-\gamma}\right),\label{muI} 
\end{eqnarray}
respectively. All these multipliers disappear at the no vaccination limit $\omega=0$.

Hence, it is interesting to note that the non-zero vaccination rate, $\omega$, enters the approximate algebraic expressions for the positions and magnitudes of the first peaks for the time evolutions $E(t)$ and $I(t)$ in two places. The first one is the explicit dependence of the critical identification rate, $\alpha_c$ on $\omega$; and the second one is the presence of additional, $\omega$-dependent factors, in equations~(\ref{tauE})--(\ref{muI}). The other terms are absolutely the same as for the no vaccination case, given by equations~(\ref{A})--(\ref{q}). As judged from the positions and magnitudes of the first peak evaluated from equations~(\ref{tmaxEv})--(\ref{alphaIc}), that are shown in figure~\ref{vacc_EI_evol} via dashed lines, the approximations are rather good and match well the time evolution oscillations obtained via numeric integration.

\section{Cellular automaton simulations}\label {IV}

The compartmental \textit{SEIRS} model, considered in sections \ref{II} and \ref{III}, has a number of limitations. One of them is the assumption of perfect miscibility between the individuals comprising the population. It is assumed that during a discrete time step, any susceptible individual can meet any unidentified infective individual and contract the disease with the constant probability $\beta$. As discussed above, it is a composite probability that includes: probability for two individuals to meet, probability that the infected individual spreads virus around, and probability that the susceptible individual is infected. In real life, all three may be quite independent and subject to certain distributions. The approach with a separate compartment of exposed individuals \cite{Diagne2021, LaJoie2022, Jdid2024, Juga2025} splits the meeting and disease transmission events into two. However, even in this case, the model does not account for the effects of social contacts, e.g., in the form of communication networks \cite{Holovatch2017,Amati2018,Blavatska2022} and, on a larger scale, on geography-based population arrangement. This limitation is caused by the mean-field nature of any compartmental model.

To take into account the spatial arrangement of individuals within the \textit{SEIRS} model as defined in figure~\ref{Model}, we consider their simplest 2D arrangement: a simple square lattice with a neighbourhood size for each individual, given by the number $q$ of its neighbours. The setup is similar to that being discussed for the case of a simpler \emph{SIS} model in reference~\cite{Ilnytskyi2016}. In particular, the vertex set of a graph is $\mathbb{Z}^2$, and each individual is characterized by a neighbourhood radius $R_q>0$, such that the neighbourhood of a given $k \in \mathbb{Z}^2$ is defined as: $k' \sim k$ whenever $|k'-k|\leqslant R_q$. The choice of $R_q$ is made to obtain the neighbourhood size equal to the previously chosen neighbourhood size $q$. Each $k$ is associated with an individual, that is characterized by its state $s_k$, which takes any value from a set of $\{s,e,i,r\}$, the letters correspond to the susceptible, unidentified infected, identified infected and recovered states, respectively. The evolution of the system of $N$ such individuals is driven by the algorithm, based on transition rules between the states of each individual in the \textit{SEIRS} model, depicted schematically in figure~\ref{Model}. The principal difference with the compartmental realization of this model is that the transmission occurs locally, within the sphere of the radius $R_q$ around each infected individual. 

We consider a square lattice of $N=700^2=490\,000$ individuals which brings the model to a realistic size of a medium-sized town. The periodic boundary conditions are applied along both Cartesian axes. The set of neighbourhood sizes ranges from $q=N$ (the limit of the compartmental model) down to $q=4$ (von Neumann neighbourhood). In this way the effect of quarantine is considered explicitly, by reducing the neighbourhood size $q$, and can be partly separated from the probability to contract a disease. At each time instance, we made $N$ random choices of individuals, and their states are updated according to their current state and the states of their neighbours. This levels up the time discretization with the $\Delta t$ of the numeric integration (\ref{integr}). We employed the asynchronous realization of the cellular automaton algorithm, when the change of $s_k$ occurs immediately, and affects the rest of $N$ updating attempts performed at the same time instance.

We focus here on the time evolution of the fraction of identified infected individuals, $I(t)$, as it characterizes the load put on the medical system of an imaginary town. The strength of quarantine measures is defined by restrictions imposed on social contacts defined via the neighbourhood size $q$. It is fixed throughout the system of $N$ individuals, but other options are equally possible. For instance, $q$ can be  chosen at random from some interval, $[0,q_{\mathrm{max}}]$ \cite{Ilnytskyi2016}, or according to some specified distribution. In the latter case, one can mimic certain network structures \cite{Holovatch2017,Amati2018,Blavatska2022} in an implicit way.

\begin{figure}[!h]
	\begin{center}
		\includegraphics[clip,width=14cm,angle=0]{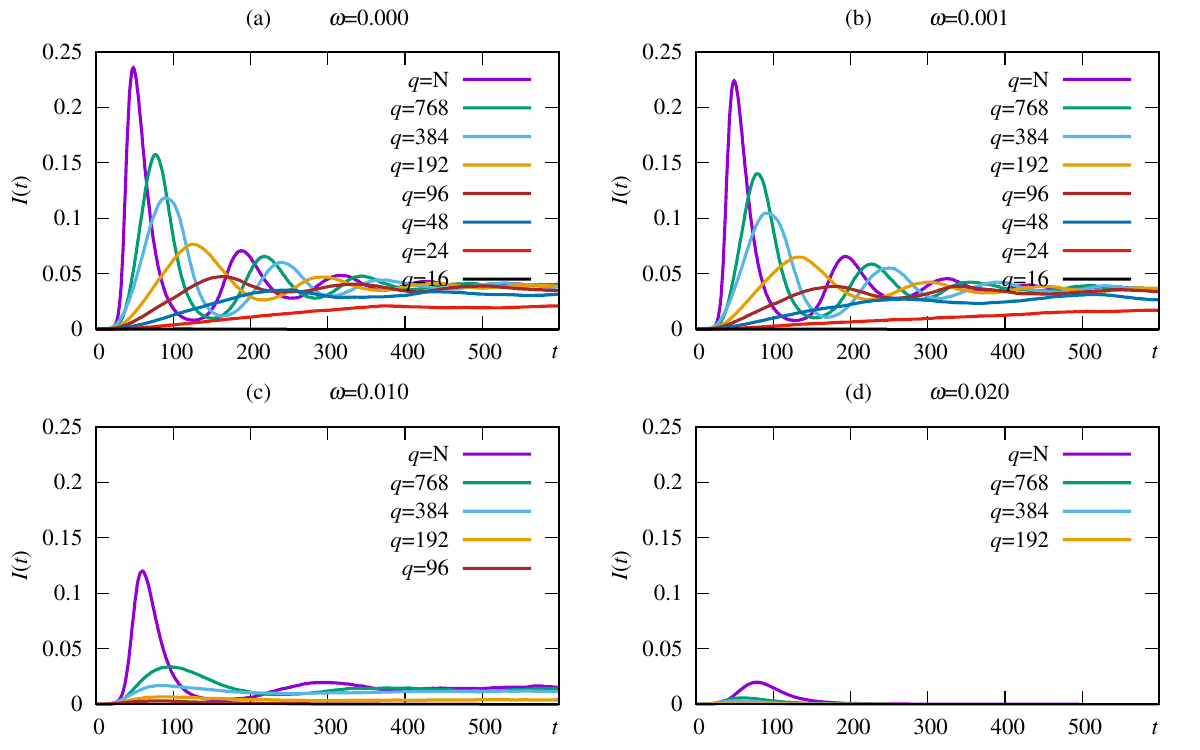}
		\caption{\label{CA_I_evol}(Colour online) The effect of vaccination rate $\omega$ and of the quarantine measures (via the neighbourhood size $q$) on time evolution of the fraction $I(t)$. The case of $\beta=0.5$, $\alpha=0.1$ is shown for the system of $N=490\,000$ individuals on a square lattice with the system evolution driven via cellular automaton algorithm. Respective values for $\omega$ for each frame, (a)--(d), are indicated in the figure.}
	\end{center}
\end{figure}

Figure~\ref{CA_I_evol} shows the time evolution $I(t)$ for four cases: no ($\omega=0$), slow ($\omega=0.001$), fast ($\omega=0.010$) and ultrafast ($\omega=0.020$) vaccination at fixed $\beta=0.5$, $\alpha=0.1$ as shown in the frames (a)--(d), respectively. For each of these cases we consider a wide set of neighbourhood sizes, $q$, as indicated in the plots. For the no vaccination case, the quarantine measures resulted in two effects: the first is a shift of a first and consequent peaks to longer times, and the second is lowering their maximum values. Let us note that the stationary state (at least, its estimate at large $t\sim 600$) is still a nonzero value $I(600)>0$, until very strict quarantine measures ($q=16$) are undertaken. Of course, such measures affect the society strongly, both economically and in a mental way. The situation is quite similar for the slow vaccination case ($\omega=0.001$), where small shifts of all the peaks to the right and small decrease of their maxima are observed. The behaviour of $I(t)$ changes drastically for the fast vaccination case ($\omega=0.010$), where one observes not only an essential lowering of all peaks, but also a decrease of the $I(600)$ value. Suppression of the epidemic outbreak can be achieved at much relaxed quarantine measures, the neighbourhood size of $q=768$ instead of $q=48$ for the no vaccination case. This tendency strengthens further at the yet further increase of vaccination rate, $\omega=0.020$. In this case, even for no quarantine measures, the outbreak does not progress beyond $I\approx 0.02$ for the current choice of the $\beta$ and $\alpha$ rates. Similar results are obtained for the other choices of $\beta$ and $\alpha$, which are not shown for the sake of brevity.

\begin{figure}[!h]
	\begin{center}
		\includegraphics[clip,width=10cm,angle=0]{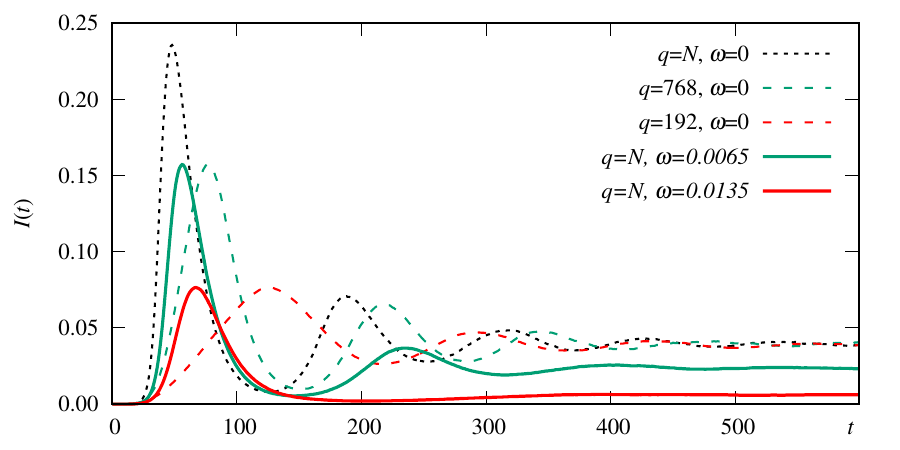}
		\caption{\label{CA_I_match}(Colour online) An attempt to match the effect of vaccination (different $\omega$ rates) and that of the quarantine measures (different neighbourhood sizes $q$), as seen on time evolution of the fraction $I(t)$. The case of $\beta=0.5$, $\alpha=0.1$ is shown for the system of $N=490\,000$ individuals on a square lattice with the system evolution driven via cellular automaton algorithm.}
	\end{center}
\end{figure}

As discussed in section~\ref{II}, real life demands a good balancing between quarantine and vaccination measures in terms of minimizing economic burden. To this end, we tried to match the cases, when only one, either vaccination, or quarantine, measure is implemented, for the same set of $\beta=0.5$ and $\alpha=0.1$ rates. The result is shown in figure~\ref{CA_I_match}, where we matched the height of the first peak in both cases. In this way, the vaccination rate $\omega=0.0065$ can be associated with the quarantine measures of $q=768$, whereas the vaccination rate $\omega=0.0135$ with $q=192$, respectively. However, one should remark that equivalent vaccination rate, found in this way, does not shift the position of the first peak (observed by implementing equivalent quarantine measures), and leads to a much stronger reduction of the height of the second peak as compared to equivalent quarantine measures. Therefore, the exact match between two measures is impossible, but, nevertheless, the results given in figures~\ref{CA_I_evol} and \ref{CA_I_match} provide some trends that may help in balancing between these two measures.

\section{Conclusions}\label {V}

We propose here the \textit{SEIRS} epidemiology model, which takes into consideration principal features of the COVID-19 pandemic: abundance of unidentified (asymptomatic or with mild symptoms) infected individuals, limited time of immunity and a possibility of vaccination. It assumes the presence of four states for individuals: susceptible; infected/infective but not yet identified; infective/infective identified and isolated; and recovered but with a temporal immunity. Vaccination changes the state from susceptible straight to recovered. The model involves: transmission rate $\beta$, identification rate $\alpha$, vaccination rate $\omega$, that can be varied, and fixed curing rate $\gamma$ and loss of immunity rate $\varphi$ based on an average statistics.

For the case of a compartmental realization of this model, we found two stationary states (fixed points) for the set of differential equations of the model: the disease-free and the endemic one. They exist in their respective restricted regions of the parameter space due to  the limitations for all fractions to stay positive and do not exceed $1$. The linear stability analysis indicates the stability of both fixed points within their respective regions. The expression of the basic reproductive number enables us to discuss the ways to lower the number of infected individuals in a stationary state via changing the model parameters. This can be achieved by lowering the contact rate $\beta$ and/or by the increase of the identification $\alpha$ and vaccination $\omega$ rates. However, if $\beta$ equals or exceeds the critical value $\beta_c$, the disease-free fixed point cannot be achieved at any combination of $\alpha$ and $\delta$. The expression for the basic reproductive number enables prediction of the most balanced approach to bring a pandemic down in economic terms, if the rates can be estimated from the real-life statistics.

Numeric integration is employed to obtain the time evolution for the fractions of infected/infective individuals of both types. The results for the positions and heights of the first pandemic peak, obtained numerically, are fitted to simple algebraic forms. These are based on observations that the positions of the peaks diverge and their heights turn into zero, when the identification rate $\alpha$ approaches the critical value $\alpha_c$. The obtained algebraic expressions provide means for simple estimates for the first peak positions and height depending on the set of values for model rates.

To separate the effect of quarantine measures from infective rate, we performed computer simulations of a lattice-based realization for this model using the cellular automaton algorithm. We found, that the quarantine measures delay and lower the first and subsequent peaks for the fraction of infected individuals, but do not lead to the complete elimination of the disease at long times, unless extremely restrictive quarantine measures are undertaken. By contrast, vaccination both lowers the peak magnitude and reduces the fraction of infected/infective individuals at long times. The attempt is made to match both quarantine and vaccination measures aimed at a balanced solution for effective suppression of the pandemic.    

\newpage   

\section*{Acknowledgements}

This work was supported by the National Research Foundation of Ukraine (Project No. 2020.01/0338).
The computer simulations have been performed on the computing cluster of the Yukhnovskii Institute for Condensed Matter Physics of NAS of Ukraine (Lviv, Ukraine).

\bibliographystyle{cmpj}
\bibliography{SEIR_covid}

\newpage
\ukrainianpart

\title{Моделювання епідемій SARS-CoV-2 за допомогою компартментної та комірково-автоматної моделей \textit{SEIRS}, що враховують ефекти скінченного імунітету та вакцинації}
%
%
\author{Я. Ільницький\refaddr{label1,label2},
	Т. Пацаган\refaddr{label1,label2}
}
\addresses{
	\addr{label1} Інститут фізики конденсованих систем
	імені І.Р. Юхновського НАН України, вул.~Свєнціцького~1, 79011 Львів,  Україна
	\addr{label2} Інститут прикладної математики та фундаментальних наук, Національний університет ``Львівська політехніка'', вул. С. Бандери 12, 79013 Львів, Україна 
}

\makeukrtitle

\begin{abstract}
	\tolerance=3000%
	Ми розглядаємо епідеміологічну модель \textit{SEIRS}, що враховує такі особливості спалаху COVID-19 як: наявність значної кількості неідентифікованих хворих, скінченний час імунітету та можливість вакцинації. Динаміка поширення пандемії контролюється соціальним дистанціюванням, інтенсивністю ідентифікації інфікованих індивідів, а також вакцинацією популяції. Для компартментної версії цієї моделі отримані стійкі стаціонарні стани: з повним подоланням пандемії та ендемічний стан. Базове репродуктивне число аналізується з урахуванням балансу карантинних та вакцинаційних заходів. Отримані чисельно положення та висота першого піку спалаху апроксимовані простими у використанні алгебраїчними формами. Граткова реалізація цієї моделі досліджена за допомогою алгоритму асинхронного коміркового автомату, що дало можливість вивчити вплив соціального дистанціювання безпосередньо: через вибір розміру сусідства агентів моделі. Зроблено спробу зіставити ефекти карантину та вакцинації та їх балансування.
	\keywords компартментні моделі, комірковий автомат, епідемії
	
\end{abstract}

\lastpage
\end{document}